\documentclass[conference]{IEEEtran}
\IEEEoverridecommandlockouts

\usepackage{cite}
\usepackage{amsmath,amssymb,amsfonts}
\usepackage{algorithmic}
\usepackage{graphicx}
\usepackage{textcomp}
\usepackage{indentfirst}
\usepackage{color, colortbl}
\usepackage[table, svgnames, dvipsnames]{xcolor}
\usepackage{caption}
\usepackage[english]{babel}
\usepackage{wrapfig}
\usepackage{enumitem}
\usepackage{subcaption}
\usepackage{xfrac}
\usepackage{diagbox}
\usepackage{array}
\usepackage{multirow}
\usepackage{arydshln}
\usepackage{tcolorbox}
\usepackage{mathtools}
\usepackage{geometry}

\makeatletter
\def\ps@IEEEtitlepagestyle{%
  \def\@oddfoot{\mycopyrightnotice}%
  \def\@oddhead{\hbox{}\@IEEEheaderstyle\leftmark\hfil\thepage}\relax
  \def\@evenhead{\@IEEEheaderstyle\thepage\hfil\leftmark\hbox{}}\relax
  \def\@evenfoot{}%
}
\def\mycopyrightnotice{%
  \begin{minipage}{\textwidth}
  \centering \scriptsize
  \copyright~2021 IEEE.  Personal use of this material is permitted.  Permission from IEEE must be obtained for all other uses, in any current or future media, including reprinting/republishing this material for advertising or promotional purposes, creating new collective works, for resale or redistribution to servers or lists, or reuse of any copyrighted component of this work in other works.
  \end{minipage}
}
\makeatother

\DeclareMathOperator{\EX}{\mathbb{E}}

\def\BibTeX{{\rm B\kern-.05em{\sc i\kern-.025em b}\kern-.08em
    T\kern-.1667em\lower.7ex\hbox{E}\kern-.125emX}}
\begin{document}

\title{Detecting Braess Routes: an Algorithm Accounting for Queuing Delays With an Extended Graph\\
\thanks{This project was sponsored in part by the joint NSF-DOT CPS grant CNS-1545116 - Traffic Operating System for Smart Cities.}
}

\author{\IEEEauthorblockN{Mikhail Burov}
\IEEEauthorblockA{
\textit{UC Berkeley}\\
mikaburov@berkeley.edu}
\and
\IEEEauthorblockN{Can Kizilkale}
\IEEEauthorblockA{
\textit{UC Berkeley and LBNL}\\
cankizilkale@berkeley.edu}
\and
\IEEEauthorblockN{Alexander Kurzhanskiy}
\IEEEauthorblockA{
\textit{UC Berkeley}\\
akurzhan@berkeley.edu }
\and
\IEEEauthorblockN{Murat Arcak}
\IEEEauthorblockA{
\textit{UC Berkeley}\\
arcak@berkeley.edu}
}

\maketitle

\begin{abstract}
The Braess paradox is a counter-intuitive phenomenon whereby adding roads to a network results in higher travel time at equilibrium. In this paper we present an algorithm to detect the occurrence of this paradox in real-world networks with the help of an improved graph representation accounting for queues. The addition of queues to the network representation enables a closer match with real data. Moreover, we search for {\it routes} causing this phenomenon (`Braess routes') rather than links, and advocate removing such routes virtually from navigation systems so that the associated links can continue to serve other routes. Our algorithm relies on a convex optimization problem utilizing Beckmann potentials for road links as well as queues, and results in a route reconfiguration with reduced delay. We assume the availability of historical data to build the optimization model. We also assume the existence of a centralized navigation system to manage the routing options and remove the Braess routes. The theoretical solution demonstrates up to 12\% delay reduction in a network from Montgomery  County,  Maryland. We validate the improvement with simulations.

\end{abstract}


\section{Introduction}

Road networks suffer from various types of equilibrium inefficiency due to selfish routing. Of particular interest is the Braess paradox, a counter-intuitive phenomenon describing scenarios in which building new road links results in higher traffic delays at equilibrium. Since its introduction in 1968 in \cite{braess1968paradoxon}, the Braess paradox has been studied extensively to find efficient ways to predict, detect and prevent its existence.

Early results, such as \cite{murchland1970braess}, \cite{leblanc1975algorithm}, \cite{fisk1979more}, \cite{stewart1980equilibrium}, \cite{frank1981braess}, focus on the classic diamond-shaped, four-node network with a single OD-pair. A later study \cite{steinberg} extends the analysis to general traffic networks to predict the occurrence of the paradox; however, the applicability of the results is limited by a restrictive assumption that all routes with non-zero flows in the original network are also utilized after the addition of a road link. 
Moreover, \cite{steinberg} as well as related theoretical papers \cite{dafermos1984some} and \cite{taguchi1982braess} consider the special case when exactly one road link or route is built or removed from the network.

Another approach to anticipating the Braess paradox is a network topology analysis. The study \cite{Milchtaich} shows that a two-terminal network is immune to Braess paradox if and only if it is series-parallel. In addition, the paper extends the result to account for any Pareto inefficiency in a two-terminal network. References \cite{cenciarelli2016graph} and \cite{chen2015excluding} extend the characterization from undirected graphs to directed graphs and allow for multiple commodities. Another theoretical study, \cite{fujishige}, explores the concept of matroids to identify networks that are immune to the Braess paradox. Unfortunately, few transportation networks exhibit a matroid structure; therefore, the applicability is limited in practice. A further shortcoming of the theoretical studies mentioned above is that they present structures that are immune to the Braess paradox, but do not provide tools to modify existing networks to eliminate this paradox and improve the efficiency of the equilibrium.

The computational study \cite{ji2014detecting} proposes a mathematical programming method to detect the Braess paradox in a given network. The problem is formulated as a bi-level structure and then transformed into a single-level mixed integer program. By setting tolls to links and analysing the resulting network latency, the algorithm detects links that cause the Braess paradox and penalizes them to imitate road closure. 

Instead of links, in our paper we search for `Braess routes' (routes that cause the Braess paradox) with a greedy optimization algorithm and remove them sequentially from the navigation system. This leaves the drivers with a subset of routes that are immune to the paradox, which they are free to choose from. We believe that removing routes is advantageous over removing links, as removing a link adversely affects all other routes going through this link. Among other scenarios, our approach enables removing through-traffic from residential roads, while allowing them to continue to serve the residents. One shortcoming of the route removal approach, however, is that customers might simply switch to another navigation system if their freedom of choice is limited. Further research is needed to address this issue, such as splitting populations into ``selfish" and ``altruistic" to model different levels of cooperation or creating a system of benefits to encourage drivers' participation. 

Unlike other methods for detecting and eliminating the Braess paradox, our model accounts for intersection structures and queues. A more accurate graph representation  achieved by fitting functions for link delays and queue delays from data allows us to make theoretical methods applicable \newgeometry{letterpaper, top=54pt,bottom=54pt,right=54pt,left=54pt} \noindent to real-world networks. We validate our approach on an extended graph model of a road network from North Bethesda, Montgomery  County, Maryland. We slightly modified the geometry of this network with additional links to be able to test a bigger number of OD-pairs and routes. We were able to demonstrate up to 12\% delay reduction on this extended network. Despite this improvement, we do not claim the proposed route removal strategy is optimal. Indeed, as shown in \cite{roughgarden}, the problem of finding the optimal subnetwork is NP-hard. Instead we trade optimality with computational tractability and applicability to real-world networks.

\section{Network Specification}

We consider a network with several routes available to each OD-pair. Every route is represented by a set of consecutive links connecting the origin to the destination. 

\subsection{Link delay function}

We estimate the time delay that a vehicle experiences on each link with a Bureau of Public Roads (BPR) function \cite{united1964traffic}:
\begin{equation}
    \Phi(z) = t_0\left(1 + a\left(\frac{z}{cap}\right)^b\right),
\end{equation}
where 

$t_0$ is the free-flow time  ($t_0 = \frac{link-length}{free-flow-speed}$),

$a$ and $b$ are parameters that depend on the link's properties,

$cap$ is the link throughput capacity (in $\frac{veh}{hour}$).

\begin{figure}[t!]
  	\includegraphics[width=1.0\linewidth]{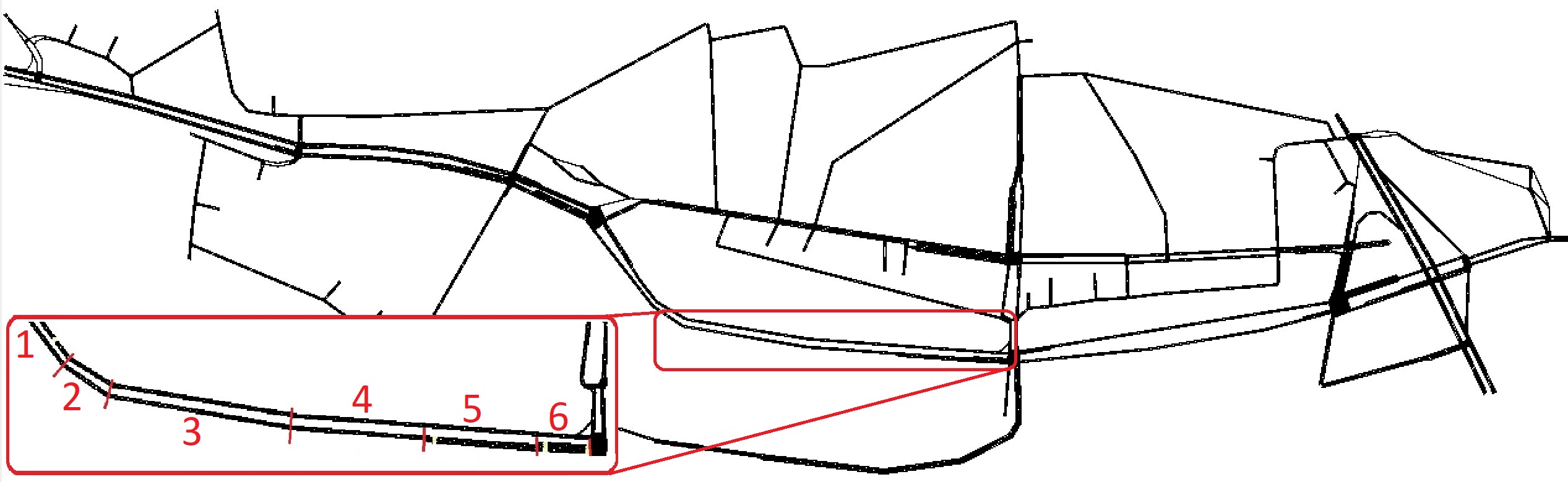}
  	\caption{Montgomery County network, slightly modified with additional links to test a bigger number of OD-pairs and routes. The red box shows how one edge is represented with a series of links, as discussed in Appendix \ref{app_delay}.}
  	\label{Mont}
\end{figure}

To compute the values of parameters $a$ and $b$ for a particular link, we followed the method described in \cite{kucharski2017estimating}, which requires speed-density data points for a curve-fitting algorithm that outputs appropriate parameters. We used Simulation of Urban Mobility (SUMO) open-source software to build a test case based on the modified North Bethesda, Montgomery County, Maryland network around the intersections of Montrose Rd and Montrose Pkwy (Fig. \ref{Mont}). In the absence of historical data, we generated traffic data from the simulation environment and estimated the delay functions. The derivation is detailed in Appendix \ref{app_delay}. 

\subsection{Queue delay estimation}

\begin{figure}[h!]
  	\includegraphics[width=1.0\linewidth]{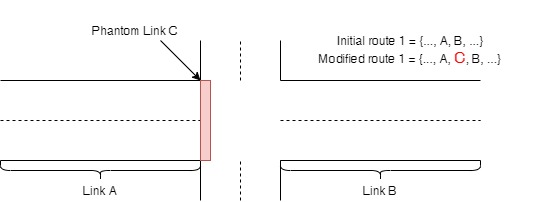}
  	\caption{``Phantom" link insertion.}
  	\label{phan}
\end{figure}

Link delay functions do not account for the queues, which accumulate when the link throughput capacity is insufficient for the incoming flow. To address this issue and model queue delay we introduce ``phantom" links, i.e. links that have no physical analogue in the real-world or simulation, but exist solely to account for additional delay related to queues. We insert ``phantom" links into routes between consecutive edges incoming to intersections and edges leaving the same intersections (Fig. \ref{phan}).   

The analysis in Appendix \ref{app_queue} yields queue delay function:

\begin{equation}
    \Phi_q(z) = \begin{cases}
                d_0 & \text{if $z < s$} \\
                \alpha z + \beta & \text{otherwise,}
             \end{cases}
    \label{queue_delay}
\end{equation}
where 

$d_0$ is expected constant delay due to intersection structure,

$\alpha$ and $\beta$ are parameters that depend on the link's properties,

$s$ is the link's saturation rate.

The proposed function is a continuous non-decreasing non-negative piece-wise linear function. Therefore, it can be readily used in our route detection algorithm.

\section{Algorithm}

\subsection{Wardrop Equilibrium computation}

To compute the Wardrop Equilibrium state we use a convex optimization problem that utilizes Beckmann potentials \cite{beckmann1956studies} for delay functions. The objective function of the optimization problem to minimize is the sum of Beckmann potentials, i.e. integrals of delay functions, across all links:

\begin{equation}
	  \displaystyle\sum_{i \in L} \phi_i(z_i)
	  \label{obj}
\end{equation}
where

$L$ is the set of links in the network,

$\phi_i$ is the Beckmann potential of the link $i$ ($\phi_i^{'}= \Phi_i$),

$\Phi_i$ is the delay function of the link $i$,

$z_i$ is the flow on the link $i$ ($z_i = (R^{T}x)_i$),

$x$ is the vector of route flows,

$R$ is the routing matrix defined as

\begin{equation}
    R_{ij} = \begin{cases}
                1, \text{ if the route $i$ goes through the link $j$} \\
                0, \text{ otherwise.}
             \end{cases}
    \label{routing}
\end{equation}

The first constraint ensures non-negative flows on routes:

\begin{equation}
    x^k_j \geq 0,\ \forall j \in P^k; \forall k \in O,
    \label{non_neg}
\end{equation}
where 

$P^k$ is the set of routes corresponding to the OD-pair $k$,

$O$ is the set of OD-pairs.

The second constraint guarantees that flows on routes corresponding to the same OD-pair sum up to the demand for that pair:

\begin{equation}
    \displaystyle\sum_{j \in P^k} x^k_j = d_k,\ \forall k \in O,
    \label{eq_dem}
\end{equation}
where
$d_k$ is the demand on the OD-pair $k$.

Therefore, the problem is:

\begin{equation} \label{optprob}
	 \underset{x}{\text{min}} \ 
	  (\ref{obj})  \
	 \text{subject to} \ (\ref{non_neg}) - (\ref{eq_dem}). 
\end{equation}

The solution of the problem (\ref{optprob}) is the vector $x^*$ of route flows at equilibrium, which results in the total network delay:

\begin{equation}
    Y = x^{*T}R \begin{bmatrix}
                    \Phi_1(z_1^*) \\
                    ... \\
                    \Phi_m(z_m^*)
                \end{bmatrix}_{|z_i^* = (R^Tx^*)_i}.
\end{equation}

\subsection{Elimination procedure}

The objective of our algorithm is to find subsets of links/routes such that the remaining network has reduced latency. To achieve this we remove links and routes that cause the Braess paradox, which we refer to as `Braess' links/routes, from the original system. Removing a route implies removing it from the set of options suggested by the navigation system (e.g., Google Maps); removing a link implies either physically closing down the road segment or reducing its capacity with tolls or signaling. As explained in the Introduction, route removal is preferable in practice; however, for the completeness of the study we discuss several approaches for route and link elimination, as they are easily obtained from the equilibrium computation method of the previous section. An important constraint in these approaches is to keep the connectivity of the network  unchanged, i.e. every OD-pair from the original network must retain at least one route in the reduced network.

In each approach we attribute values to the links/routes with the help of the optimization problem (\ref{optprob}). First we obtain the equilibrium network delay, $Y$, for the original system. Then, we tentatively remove a link/route from the network (remove corresponding rows and columns from the routing matrix (\ref{routing})) and solve (\ref{optprob}) again, deriving the new equilibrium network delay, $Y_{new}$. Having both the original and the new network delays, the value $V$ of the removed link/route is:
\begin{equation}
    V = Y_{new} - Y.
\end{equation}
Links/routes with negative values are associated with the Braess paradox, because removing them reduces latency. 

\smallskip

\noindent
 \textbf{Greedy Single Link Removal}
    \begin{enumerate}
        \item Find the link with the smallest value $V_{min}$.
        \item If $V_{min} < 0$, remove the link and revert to step (1).
        \item Otherwise, terminate.
    \end{enumerate}
\noindent
As stated earlier, removing a link affects the entire set of routes going through this link, which compromises the effectiveness of the algorithm. Moreover, this approach is slower than some algorithms discussed further. 

\smallskip

\noindent
 \textbf{Link Combination Removal}
    \begin{enumerate}
        \item Compute the values of all possible combinations of links and find the one with the smallest value $V_{min}$.
        \item If $V_{min} < 0$, remove the corresponding combination.
        \item Otherwise, the network is Braess paradox-free.
    \end{enumerate} 
\noindent
Since the number of link combinations grows exponentially with the number of links, this method is computationally expensive and does not scale well to large networks.
 \smallskip

\noindent
  \textbf{Link-Route Combination Removal}
    This approach modifies the first method and tries to address the issue with subset of routes elimination. Instead of removing links completely,
    \begin{enumerate}
        \item For each link find the subset of routes utilizing this link that results in the minimal network delay.
        \item Remove the routes that were not present in at least one optimal configuration.
    \end{enumerate}
\noindent
Unlike previous algorithms, this method removes routes, but does it indirectly by working with link-route combinations. 
\smallskip

\noindent
 \textbf{Greedy Single Route Removal}
    \begin{enumerate}
        \item Find the route with the smallest value $V_{min}$.
        \item If $V_{min} < 0$, remove the route and revert to step (1).
        \item Otherwise, terminate.
    \end{enumerate}
This approach, unlike the first method,  deals with routes directly, which is faster since in practice the number of commonly used routes is smaller than the number of links. An OD-pair can potentially have a large number of possible routes; however the dominant part of the drivers uses the a very limited subset of those routes. It also allows to identify the occurrence of the Braess paradox after the first iteration: {if the route with the minimal value has zero initial flow at equilibrium, then the network is Braess paradox-free.}
\smallskip

\noindent
 \textbf{Route Combination Removal}
    \begin{enumerate}
        \item Compute the values of all possible combinations of routes and find the one with the smallest value $V_{min}$.
        \item If $V_{min} < 0$, remove the corresponding combination.
        \item Otherwise, the network is paradox-free.
    \end{enumerate}
   This approach is similar to the second approach and has the same limitation due to computational complexity.

All five approaches were tested to identify the fastest and most accurate method for further implementation and verification. The improvements achieved on the sample network were comparable. Therefore, we report below the results for the fourth method (Greedy Single Route Removal) due to its computational speed and advantages in implementation.

\section{Simulation}

Series of simulations were conducted to demonstrate the improvement, evaluate the prediction accuracy and estimate the computational precision of our algorithm when applied to physical networks. The testing  was designed to model the behavior of a real-world traffic system, therefore a complex structure of OD-pairs and corresponding routes was built upon (Fig. \ref{Mont}). The parameters of the simulation model are presented in Table \ref{sim_param}.

\begin{table}[htb]
    \caption{Simulation parameters.}
	\center
	\begin{tabular}{ |p{4.2cm}||p{1.3cm}|  }
 		\hline
 		Parameters & Values \\
 		\hline \hline
 		Number of OD-pairs & 12 \\
		\hline
 		Number of routes per OD-pair & 1-5 \\
		\hline
 		Total number of routes & 38 \\
 		\hline
 		Number of simulated vehicles & 500 - 3200 \\
 		\hline
	\end{tabular}
    \label{sim_param}
\end{table}

A side benefit of the route removal approach is that it may help small side-roads in residential areas that usually suffer from congestion due to drivers attempting to avoid busy freeways. In the era of Google maps, Waze and other navigation systems, which try to discover short-cuts and guide vehicles through neighborhoods to free up main roads, residents of these regions face busy traffic consequences and spend significantly more time than usual on short trips \cite{Mac19}.

The testing procedure consists of the following steps:

\begin{enumerate}
    \item Solve the problem (\ref{optprob}) for the original network to obtain the equilibrium flow distribution, $x^*$, and equilibrium network delay, $Y$.
    
    \item Derive the route configuration resulting in minimal delay via route elimination algorithm and reduce the network.
    
    \item Solve the problem (\ref{optprob}) for the reduced network to obtain the new equilibrium flow, $x^*_{new}$, and network delay, $Y_{new}$. The theoretical delay reduction is: $I_{th} = \frac{Y - Y_{new}}{Y}$.
    
    \item Feed $x^*$ into the SUMO to obtain the simulation network delay, $Y^{sim}$, as a sum of travel times of all vehicles.
    
    \item Feed $x^*_{new}$ into the SUMO to obtain the new simulated network delay, $Y^{sim}_{new}$. The simulation delay reduction is $I_{sim} = \frac{Y^{sim} - Y^{sim}_{new}}{Y^{sim}}$.
    
    \item Compare the theoretical and simulated delay reductions $I_{th}$ and $I_{sim}$, and estimated total delays $Y$ and $Y^{sim}$.
\end{enumerate}

Every simulation set was run with different initial demands corresponding to OD-pairs (last row of Table \ref{sim_param}). Congestion scenarios were of particular interest, because they allowed us to test the developed queue delay estimation model.

It is important to mention that our model does not account for spillbacks (full occupation of a link that causes the queue propagation to the upstream edge). Therefore, upper bounds on demands were applied to avoid spillbacks. Further research is required for an appropriate spillback representation.

\section{Results}

In this section we present the results of our algorithm for several scenarios. The first two rows in Table \ref{results} show the size of the simulated traffic. We addressed conditions associated with different times of day.

\begin{table}[htb]
    \caption{Simulation results.}
	\center
	\begin{tabular}{ |p{2.4cm}||p{1.0cm}|p{1.0cm}|p{1.0cm}|p{1.0cm}|  }
 		\hline
 		{} & Set-up 1 & Set-up 2 & Set-up 3 & Set-up 4 \\
 		\hline \hline
 		Demand & Low & Medium & High & High \\
		\hline
 		Number of vehicles & 500 & 1600 & 2600 & 3300 \\
		\hline
 		Improvement ($I_{th}$) & 0\% & 3.2\% & 11.8\% & 8.1\% \\
 		\hline
 		Improvement Diff. & 0\% & 5.3\% & 15.3\% & 10.7\% \\
 		\hline
 		Network Delay Diff. & 1.7\% & 8.1\% & 9.5\% & 9.2\% \\
 		\hline
	\end{tabular}
    \label{results}
\end{table}

We are interested in three main metrics when analyzing the efficiency of our approach, which are presented in Table \ref{results}. The first metric (Row 3) is the theoretical improvement in the network delay, i.e. total travel time saved by all vehicles after implementing our algorithm. In the low demand scenario, the network was Braess paradox-free, and no travel time reduction was achieved. For moderate and congested traffic the improvement ranges from 3.2\% to 11.8\%, which is reasonable considering the insignificant effort required to modify the network configuration.

The second metric (Row 4) shows the difference between the theoretically predicted delay improvement and the corresponding simulation result. For the free-flow set-up there was no Braess paradox and no comparison was needed. For medium and high demands, the difference varies between 5.3\% and 15.3\%. Taking into account the improvement value itself, we can conclude that the algorithm makes a relatively accurate prediction. Additionally, the theoretical solution always resulted in actual simulation improvement, i.e.,  $I_{th} > 0 \Rightarrow I_{sim} > 0$.

The third metric (Row 5) is the difference between the predicted network delay and the simulated one. According to the results, we managed to keep the deviation under 10\% for all scenarios. Furthermore, some cases demonstrated almost identical values for the network delays. The accuracy of the prediction suggests that the proposed graph representation accounting for queue delays yields a reasonable model to reflect real traffic conditions.

\section{Conclusion}

We presented a greedy optimization algorithm to detect and sequentially remove the routes that contribute to the Braess paradox. The algorithm assigns values to existing routes in terms of how much the network delay would increase if they were removed, so that routes with negative values indicate a Braess paradox. The algorithm removes the route with  the most negative value at each iteration. The resulting reduced network configuration demonstrates up to 12\% travel time reduction. The proposed extended graph representation to account for queue delays provides a major improvement in how well the theoretical predictions match the simulated ones.
We are interested in further studying the possible methods of modeling spillbacks and other congestion-related phenomena.

\appendices
\section{Link Delay Function Derivation} \label{app_delay}

\begin{figure}[h!]
    \centering
  	\begin{subfigure}{0.49\linewidth}
    	\includegraphics[width=\linewidth]{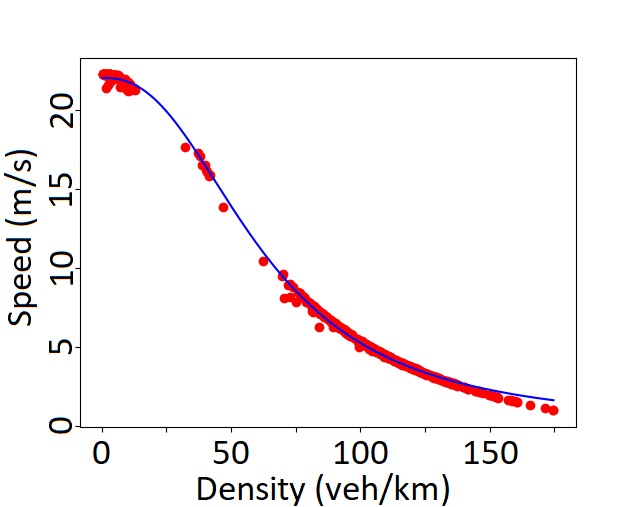}
        \caption{BPR-function fitting.}
  	\end{subfigure}
  	\begin{subfigure}{0.49\linewidth}
    	\includegraphics[width=\linewidth]{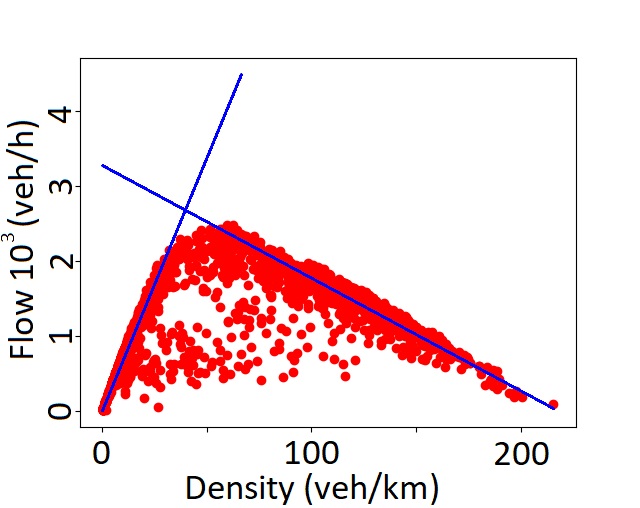}
    	\caption{Capacity estimation.}
    	\label{cap}
  	\end{subfigure}
	\caption{Sufficient data examples.}
	\label{iso_approach_suff}
\end{figure}

\begin{figure}[h!]
  	\begin{subfigure}{0.49\linewidth}
    	\includegraphics[width=\linewidth]{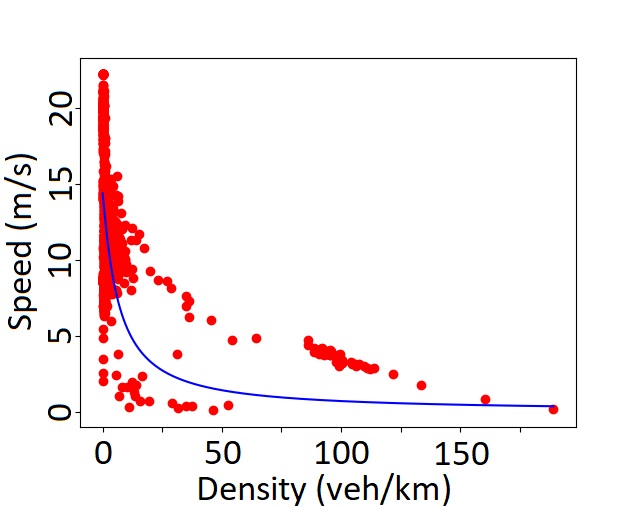}
  	\end{subfigure}
  	\begin{subfigure}{0.49\linewidth}
    	\includegraphics[width=\linewidth]{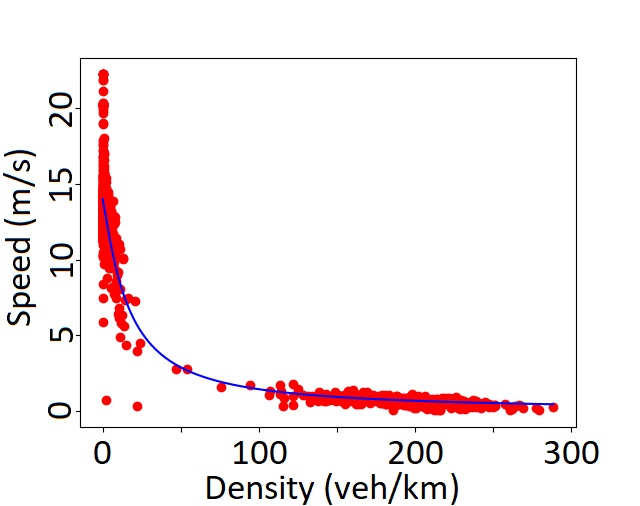}
  	\end{subfigure}
	\caption{Insufficient data examples.}
	\label{iso_approach_insuff}
\end{figure}

In this section we present a detailed link delay function derivation from empirical data points.
We treated flow data for each link individually, simulating traffic on one edge at a time. To capture a wide range of flow values, we generated a new random flow from the interval [0,3000] $\frac{veh}{h}$ every 200 seconds. The provided data were sufficient to construct BPR-functions for 90\% of links (Fig. \ref{iso_approach_suff}). Tuning simulation parameters covered additional 70\% of the remaining links (Fig. \ref{iso_approach_insuff}). For the rest of the failed edges, we used parameter values corresponding to delay functions of upstream successful links with minor modifications. We assumed these parameters are likely to have close values and, therefore, can be almost interchangeable. However, if this assumption is false the discrepancy between ground truth delay and estimated delay produced by inaccurate parameter substitution of one link is insignificant in a network scale.

The negligibility of this difference follows from the fact that road links constituting a network are relatively short in general. One edge in the graph network representation is usually displayed by a series of short connected links in the SUMO simulation network (Fig. \ref{Mont}). The travel time contribution of one link is in the order of few seconds, which is insignificantly small to deviate from the ground truth. Moreover, since the number of poorly-fitted links makes up less than 2\% of the number of all links, we can conclude that the proposed design is sufficiently accurate.

Additionally, the link capacities can be extracted from fundamental diagrams built upon the collected data (in $\frac{veh}{h}$) (Fig. \ref{cap}). The first method is to set the capacity estimate to the maximal recorded flow value. Another approach is to use a piece-wise-linear approximation to derive the capacity as a y-coordinate of the intersection of two approximation lines.

\section{Queue Delay Function Derivation} \label{app_queue}

In this section we present a detailed queue delay function derivation based on the queue formation analysis. Queues occur when the flow ($z$) on the link exceeds the saturation rate ($s$) of this link. The saturation rate is the upper bound on the number of vehicles able to leave the link within a period of time. Therefore, the queue formation depends on the difference between the inflow and the outflow on a particular link. Depending on the link type and its relative position with a specific intersection, we can distinguish several possible options for saturation rate estimation:

\begin{itemize}

    \item \textbf{If the link incomes to a signalized intersection}, its corresponding saturation rate equals to the maximum number of vehicles ($n$) able to pass the intersection on green light within one cycle normalized to one hour:
    \begin{equation}
        s = \frac{3600n}{D}
        \label{sat_tl}
    \end{equation}
    where, $D$ is the cycle duration (in seconds).
    
    \item \textbf{If there is a STOP sign at the end of the link}, its corresponding saturation rate is:
    \begin{equation}
        s = \frac{3600}{w}
        \label{sat_SL}
    \end{equation}
    where, $w$ is the delay (in seconds) a vehicle experiences when forced to stop at the STOP sign on an empty road. This value depends primarily on the speed limit on the link and has a small fluctuation from edge to edge.
    
    \item \textbf{If the link is free from any traffic regulation causing vehicles to stop or slow down}, the saturation rate equals to the link physical throughput capacity.
    
    \item \textbf{The link is the secondary link, having no priority on an unsignalized intersection}. Up to this point, the queue delay is given as $\Phi_q(z)$ which is a function of the flow on the corresponding link. In this case, however, the delay incurred by the intersection structure on the secondary link is also a function of the flow of the primary link. Although this violates the fundamental assumption of delay being a function of the corresponding link flow only, if the delay was symmetric for both the primary and secondary links, we could have still implemented this in our optimization problem and computed the equilibrium points. However, primary link sees no delay from the intersection hence the delay function is asymmetric which makes it hard if not impossible to compute equilibrium points the same way. Therefore, this scenario is not represented in our model and is subject to further research. 
    
\end{itemize}

Similar to the link delay function derivation, we need to use real-world or simulated empirical data to estimate the parameters for queue delay function. We present the detailed function computation procedure for only the first scenario, which features a link incoming to a signalized intersection. Other scenarios utilize a slightly modified approach, which we will mention at the end of this section.

Knowing the traffic light cycle length ($D$) and keeping in mind the queue size dependency on the inflow-outflow difference, we estimate the one cycle queue growth rate as:
\begin{equation}
    dq = \frac{(z - N)D}{3600},
    \label{dq}
\end{equation}
where $N$ is the throughput of the link. 

Based on the flow value, we can distinguish two scenarios:

\begin{itemize}
    \item \textbf{The flow is smaller than the saturation rate ($z < s$)}. In this case, $dq = 0$, all vehicles are able to pass the intersection and no queue is forming. To account for a potential red phase arrival, instead of setting the queue delay to zero, we choose a specific constant $d_0$, which is the expected value of a delay due to phase change:
    
    \begin{equation}
        d_0 = \EX(y) = \frac{1}{D}\displaystyle\sum_{i = 1}^{L_{red}} i = \frac{L_{red}(1+L_{red})}{2D},   
    \end{equation}
    where
    $L_{red}$ is the duration of the red phase.
    
    \item \textbf{The flow is greater than the saturation rate ($z > s$)}. In this case, $dq > 0$ and the queue is growing. To find the growth rate, the flow-throughput dependency and, thus, the saturation rate for a particular link are required. 
\end{itemize}

\begin{figure}[h!]
  	\begin{subfigure}{0.49\linewidth}
    	\includegraphics[width=\linewidth]{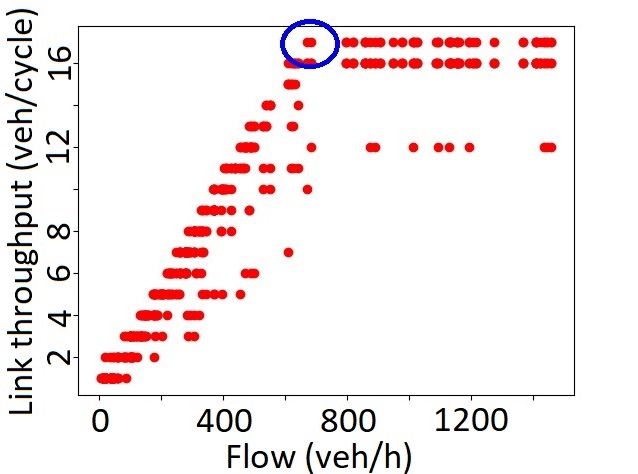}
    	\caption{Saturation rate.}
    	\label{sat}
  	\end{subfigure}
  	\begin{subfigure}{0.49\linewidth}
    	\includegraphics[width=\linewidth]{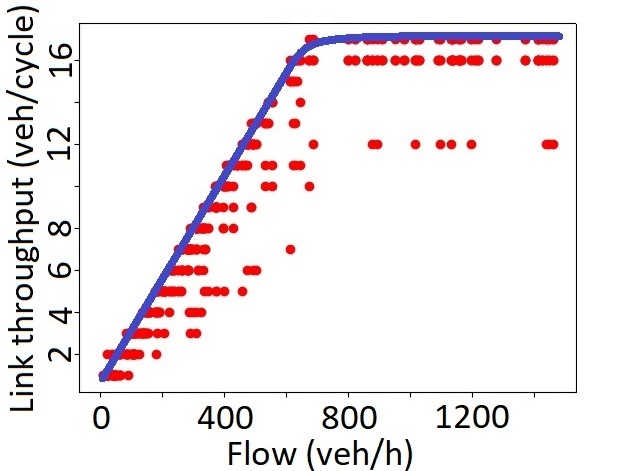}
    	\caption{Throughput function.}
    	\label{thr}
  	\end{subfigure}
	\caption{Link throughput at a signalized intersection.}
\end{figure}

Based on the queue formation model presented earlier, queue delay function depends on the saturation rate of the link. To find the one-cycle throughput capacity of the link we simulate flows of various values in ascending order for a short period of time (200 seconds) each and record the number of vehicles leaving the link (entering the intersection) within one cycle. The flow point at which the throughput linear growth stops (Fig. \ref{sat}) corresponds to the one-cycle saturation rate of the link. To obtain the value for one-hour period, scale the result by $\frac{3600}{D}$. The throughput of the link (in $\frac{veh}{h}$) is a piece-wise function of the flow which has the form:

\begin{equation}
    N = \begin{cases}
            z & \text{if $z < s$} \\
            \frac{2(z-s)}{z} + s & \text{otherwise.}
        \end{cases}
\end{equation}

For the flows smaller than the saturation rate, the throughput is a simple linear function, because all vehicles are able to leave the link.
Otherwise, the throughput equals to the sum of the saturation rate and an additional term, which never exceeds 2. This term accounts for rare scenarios with one or multiple abnormally fast-moving vehicles that manage to exceed the usual saturation rate.

\begin{figure}[h!]
    \begin{subfigure}{0.47\linewidth}
        \includegraphics[width=\linewidth]{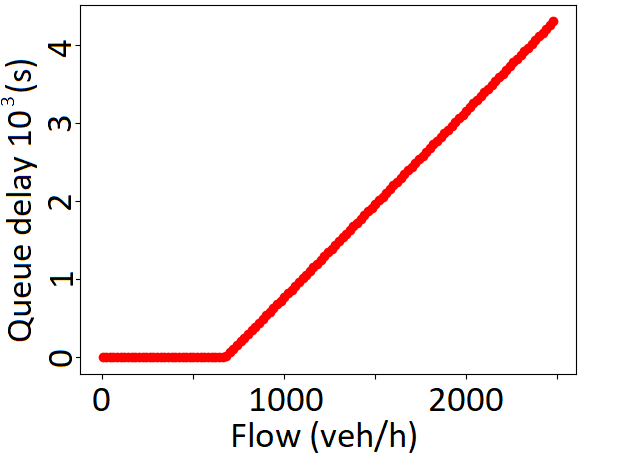}
    	\caption{Queue delay data.}
    	\label{delay}
  	\end{subfigure}
  	\begin{subfigure}{0.47\linewidth}
    	\includegraphics[width=\linewidth]{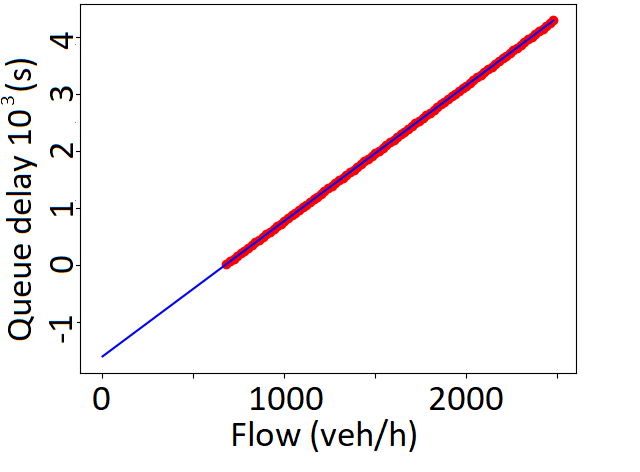}
    	\caption{Fitting delay function.}
    	\label{del_fit}
  	\end{subfigure}
	\caption{Signalised intersection flow-delay dependency.}
\end{figure}

Feeding the derived throughput function back into equation (\ref{dq}), we obtain a one-cycle gain in numbers of vehicles to the queue due to excessive flow. It is important to note that the imposed queue delay varies among vehicles within the same flow, and drivers in the head of a heavy traffic would spend significantly less time in the queue, than the ones in the tail. To avoid complications in the optimization problem formulation, an average delay can be assigned to every vehicle on the link. Plotting the computed delay results in a piece-wise linear function (Fig. \ref{delay}). 
    
The last step is to fit a linear function $\alpha z + \beta$ to the non-constant area of the graph (Fig. \ref{del_fit}) and learn the parameters $\alpha$ and $\beta$ resulting in equation (\ref{queue_delay}).

The presented approach can also be used to determine queue delay functions for links ending up with the STOP sign. Both the expected delay $d_0$ and the cycle length $D$ would be artificially set to $w$ from  equation \eqref{sat_SL}, since every vehicle without exception is required to stop at the STOP sign, which takes it exactly $w$ seconds to do.

\bibliographystyle{unsrt}
\bibliography{citing}

\end{document}